Educational game design: game elements for promoting engagement

Angela Y. He[a]

[a]Oakton High School, Sutton Road 2900, Vienna, VA 22181-2900, United States. Corresponding author. E-mail address: heangela.arts@gmail.com

This research did not receive any specific grant from funding agencies in the public, commercial, or not-for-profit sectors.

Abstract

Engagement in educational games, a recently popular academic topic, has been shown to increase learning performance, as well as a number of attitudinal factors, such as intrinsic interest and motivation. However, there is a lack of research on how games can be designed to promote engagement. This mixed methods case study aimed to discover effective game elements for promoting 17-18 year old high school students' engagement with an educational game. Using within-case and cross-case analyses and triangulated data, 10 elements emerged and were categorized into the constructs of story, gameplay, and atmosphere. Examples and connections to the literature for each element are reported. Findings implicate that educational game design for both learning and engagement is composed of educational-game specific elements, game design for solely engagement is similar for both educational and entertainment games, and a gap on educational game design technique instead of theory should be addressed to further benefit educational game development.

Educational game design: game elements for promoting engagement

## 1. Introduction

Engagement with games can be concisely defined as an investment of time, effort, and attention into a game, and significantly impacts various attitude and behavioral outcomes (Brown & Cairns, 2004; Hainey et al., 2016). In the past decades, educational video games as engaging teaching media have grown in corporate popularity and academic attention (Boyle et al., 2016; Entertainment Software Association, 2016; Vandercruysse, Vandewaetere, and Clarebout, 2012). In the 1980s, most studies on games examined their harmful effects such as induction of aggression or addiction; however, some studies noted the educational potential of games (Anderson & Ford, 1986; Selnow, 1984). For example, Cullingford, Mawdesley, and Davies (1979) found that games were observed to hold much more engagement potential than computer simulations, which had previously aided education. In the 2000s, games quickly became one of the most pervasive entertainment mediums worldwide, and academia about games, especially educational ones, increased (e.g., Aldrich, 2004; Gee, 2003; Squire, 2008). Although findings on the efficacy of educational games were sporadic and conflicting in the late 2000s (Vandercruysse, Vandewaetere, and Clarebout, 2012), findings became more steady and numerous in the 2010s, and consensus was eventually achieved on the positive effects of educational games. Among the various and positive motivational, attitudinal, behavioral, cognitive, and learning outcomes of educational games (Boyle et al. 2016; Hainey et al., 2016), researchers also found that educational games are more engaging than traditional teaching methods, such as classroom lectures (Annetta, Minogue, Holmes, & Cheng, 2009; Giannakos, 2013; Kuo, 2007; Wrzesien & Raya, 2010). The opportunity for increased engagement in educational games should be capitalized upon in research, education, and corporations; hence this study's existence.

*1.1. Defining engagement*

Previous education research (Banyte & Gadeikiene, 2015; Fredricks, Filsecker, & Lawson, 2016; Rashid & Asghar, 2016) most commonly agree that engagement is a three-dimensional construct, composed of affective engagement, behavioral engagement, and cognitive engagement. When engagement with games is defined, the definition becomes more nuanced. In Brown and Cairns's (2004) landmark paper on immersion, affective engagement is viewed as the first step to becoming engaged, with behavioral and cognitive engagement coming afterwards. In accordance with the Technology Acceptance Model (TAM), players will hold positive perceptions and beliefs about the game (affective engagement) if their initial impression of the game begets sufficient perceived usefulness; after holding positive affect and consequently becoming affectively engaged, the player invests time and attention (behavioral engagement) and effort (cognitive engagement) into learning how to play and playing the game (Brown & Cairns, 2004; Davis, 1993). Thus, engagement with games possibly holds the different dimensions of engagement in different priorities. In congruence with this line of reasoning, gaming researchers (Brown & Cairns, 2004; Hamari et al. 2016; Jennett et al., 2008; Laffan, Greaney, Barton, & Kaye, 2016) section game engagement into a hierarchy with multiple levels; the highest levels are called "flow," "immersion," or "presence" and define a state of complete absorption in the game world with little or no awareness of the external world. Conclusively, engagement with games contains a chronological, ranking component. Thus, when identifying engaging elements in educational games in this study, the level of engagement (e.g. "engrossment," "immersion") induced will be considered.

*1.1. Statement of problem*

Seeing as the efficacy of educational games has been tentatively established by past research, the gaps in educational gaming research now regard the design and development of educational games (Boyle et al., 2016). Game development often requires complex planning (e.g. game design documentation, risk assessments, prototyping) and high financial and opportunity costs, thus research that can help facilitate educational game development and possibly alleviate these barriers is beneficial (Bethke, 2003; Schell, 2014). Although frameworks for the design and development of educational games have been published (e.g., Aleven, Myers, Easterday, & Ogan, 2010; Annetta, 2010; Mitgutsch & Alvarado, 2012), many take a broad perspective, examining effective educational game design in general. Educational game design research with a more narrowed approach often focuses on elements to promote learning (Arnab et al., 2015; Garris, Ahlers, & Driskell, 2002; Guillén-Nieto & Aleson-Carbonell, 2012). Because the aim of educational games is to engage and educate (Ke, 2008; Swain, 2007), research on educational game elements promoting engagement is also of foremost importance (Boyle, et al. 2016). Engagement with a game positively affects the player' intrinsic motivation, positive emotion, germane attention, and satisfaction with the game (Arnone, Small, Chauncey, & McKenna, 2011; Brown & Cairns, 2004; Chang, Liang, Chou, & Lin, 2017; Driscoll, 2000; Hu & Hui, 2012). Via directly affecting intrinsic motivation, engagement can also indirectly improve learning performance, curiosity, exploration and experimentation (Huizenga, Admiraal, Akkerman, & Ten Dam, 2009; Hyland & Kranzow, 2011; Ryan & Deci, 2000). While engagement can affect learning, this effect does not seem to be reciprocated; it is a one-way street (Hamari et al., 2016). Given the impact of engagement and the lack of research on engaging game elements, this study places priority on designing for engagement over learning in educational games. While there is a plentitude of research on engaging game design for games in general (e.g., Laffan, Greaney, Barton, & Kaye, 2016; Pinelle, Wong, & Stach, 2008; Rigby & Ryan, 2011; Schell, 2014), this may not be applicable to educational games. Educational games have the unique, specific purpose to teach, and thus require a more narrowed approach for their game design (Annetta, 2010). This study proposes to fill this gap in educational game design.

*1.1. Purpose of study*

This study aims to advance the research on engaging game elements in educational games. Therefore, by collecting data on student engagement with this study's educational game intervention, and with educational and entertainment games in general, this study focuses on the following research question: What game elements most facilitate engagement in 17-18 year old high school students with an educational game?

## 2. Methods

*2.1. Research design*

In accordance with Cohen, Manion, and Morrison's (2013) advice, mixed methods were used to produce the most comprehensive, convincing evidence for instructional technology. Because concurring feedback from participants was needed, a multiple case study design was employed (Yin, 2013). Findings were triangulated using observations, interviews, and documents. Quantitatively, a within-group pretest-posttest comparison was collected from documents to investigate the relationship between performance and engagement.

*2.2. Participants*

Ten students from a high school in Fairfax County, Virginia participated. The school is more academically rigorous than national average and in a high-income area (US News, 2016). The sample, selected using purposeful sampling and thus the researcher's knowledge of the school's students, was intended to be representative of this study's population. The population was 17-18 years old, approximately equal proportions of males and females, mid-high income, and racially diverse. Demographics questionnaires confirmed that participants were equally distributed in gender, age, and gaming behaviors. Age was chosen because older teens are the more likely consumer of games, with the average game player being 35 (Entertainment Software Association, 2015).

Using random assignment, five students—3 female and 2 male—each were assigned to one of two groups (group A or group B).

*2.3. Materials*

*2.3.1. The two anatomy games*

Two versions (game A and game B) of *Grey Plague*, a point and click adventure game developed by the researcher, were used in this study. Group A played game A; group B played game B. Game B is free-to-play online at https://zephyo.itch.io/grey-plague. *Grey Plague*'s purpose is to engage while teaching basic concepts on human anatomy, physiology, and tuberculosis. The anatomy and physiology content conforms to the participants' high school Anatomy & Physiology curriculum. Tuberculosis content serves as a plot device and medical advanced-professional learning, i.e. learning not part of K-12 education that facilitates professional success (Mayer et al., 2013).

The learning content of *Grey Plague* specifically teaches the locations and functions of many vital organs such as the bones and those of the nervous systems and respiratory system; the definition and limitations of 3D organ printing; and the definition, infection process, and symptoms of tuberculosis.

The games' design purposefully adopted the following elements for general and educational game design (Annetta, 2010; Pinelle, Wong, & Stach, 2008; Rigby & Ryan, 2011; Schell, 2014): rules and rewarding goals, predictable controls, atmospheric music and sound effects, identity, 3-dimensional characters, intuitive user interface (Fig. 1), and visual aids for learning content (Fig. 2).

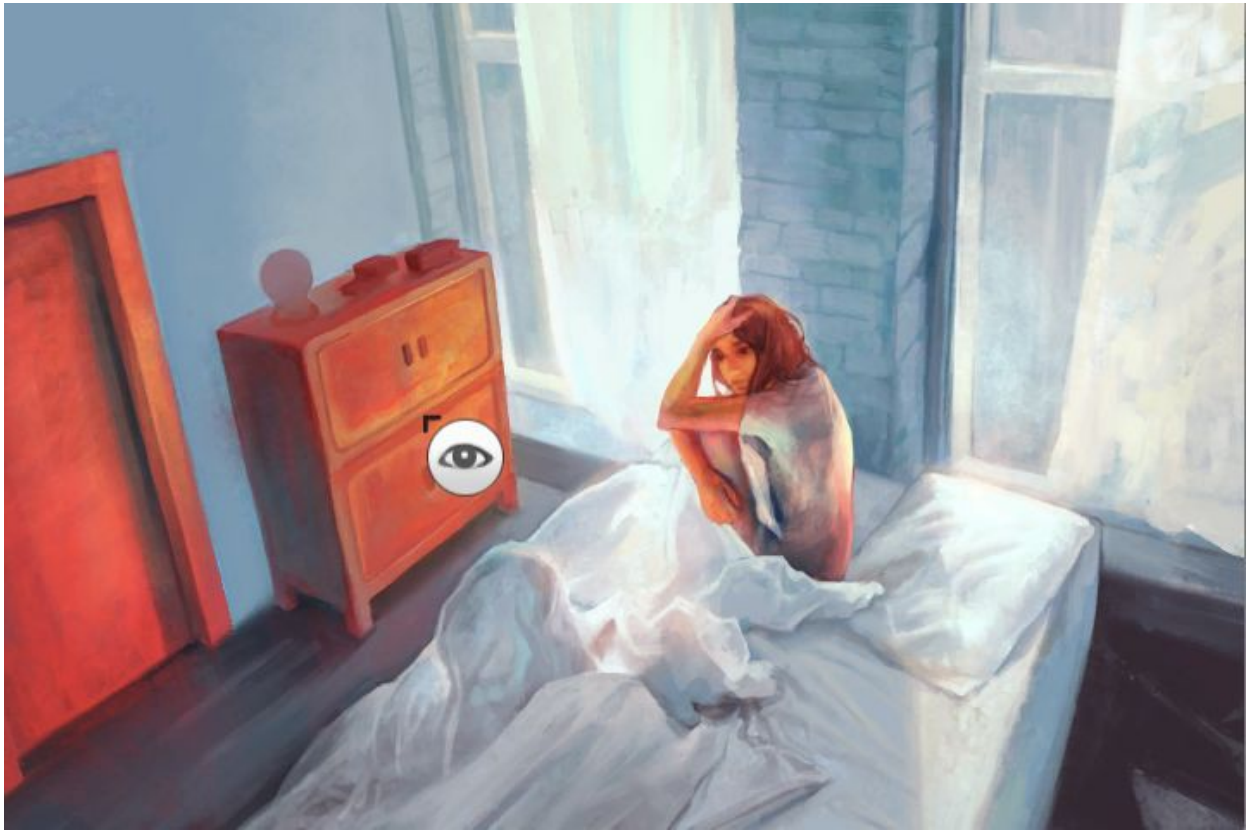

*Fig. 1*. Talking, examining, and entering actions are detected via user interface.

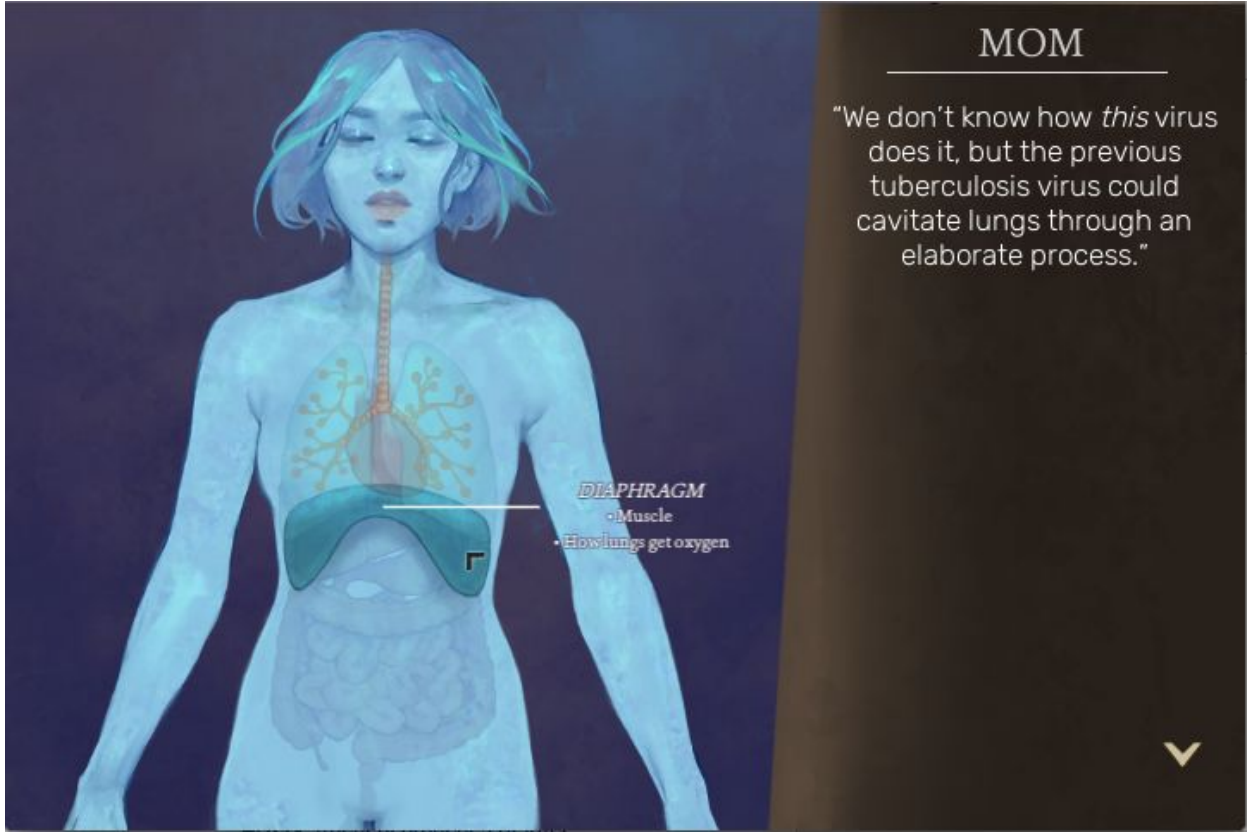

*Fig. 2*. Interactive diagrams give more information upon mouseover.

The story is based off of Vogler's (2007) *Writer's Journey*. The story takes place in the future (2070) and revolves around a main conflict between the protagonist and a deadly disease called Grey Plague, suspected to be a mutation of tuberculosis. The protagonist has the superpower to see other human's' internal anatomies and can defeat Grey Plague via this superpower and 3D printed organs, but is held back by personal anxieties; her feelings instigate interpersonal conflicts and moral dilemmas regarding her mom, grandma, best friend, and coworker.

*Grey Plague* was developed in Unity using C#, employing graphics built in Photoshop and audio sourced from Creative Commons.

2.3.1.1. Difference between game A and B

The difference between the games (hence 'differentiated section') aims to engage participants at a higher level in game B than game A. To engage participants more in game B, conscious design decisions based on existent research (Annetta, 2010; Chang, Liang, Chou, & Lin, 2017; Chen & Sun, 2012; Sanders & Cairns, 2010; Schell, 2014) were made. Specifically, in the differentiated section, the pace of the music and the amount of dynamic visuals and meaningful interaction were intentionally increased more for game B than A. The purpose of inducing different engagement levels is to diversify participants' perspectives on educational games. Thus, collected data will be more varied, and findings on game elements will have more integrity (Yin, 2013).

The differentiated section occurs when the protagonist's mother, a doctor, explains the infection process of tuberculosis to the protagonist. Participants took about 5-7 minutes playing this section. The differentiated section in game A resembled a dull visual novel; the section used interactive diagrams and animations accompanied by academically toned text of the mother's explanation (Fig. 2). In game B, the differentiated section resembles a strategy or action game, and the mother instead encourages the protagonist to imagine themselves as a tuberculosis bacterium. The protagonist then imagines and controls said bacterium, guiding it through the infection process. Game B's differentiated section is split into four levels which get progressively more difficult for the player's motor and spatial skills (Fig. 3); this difficulty, player-controlled movement, and effective multimedia motivates the player to actively solve a challenge rather than read text and look at animations, encouraging engagement.

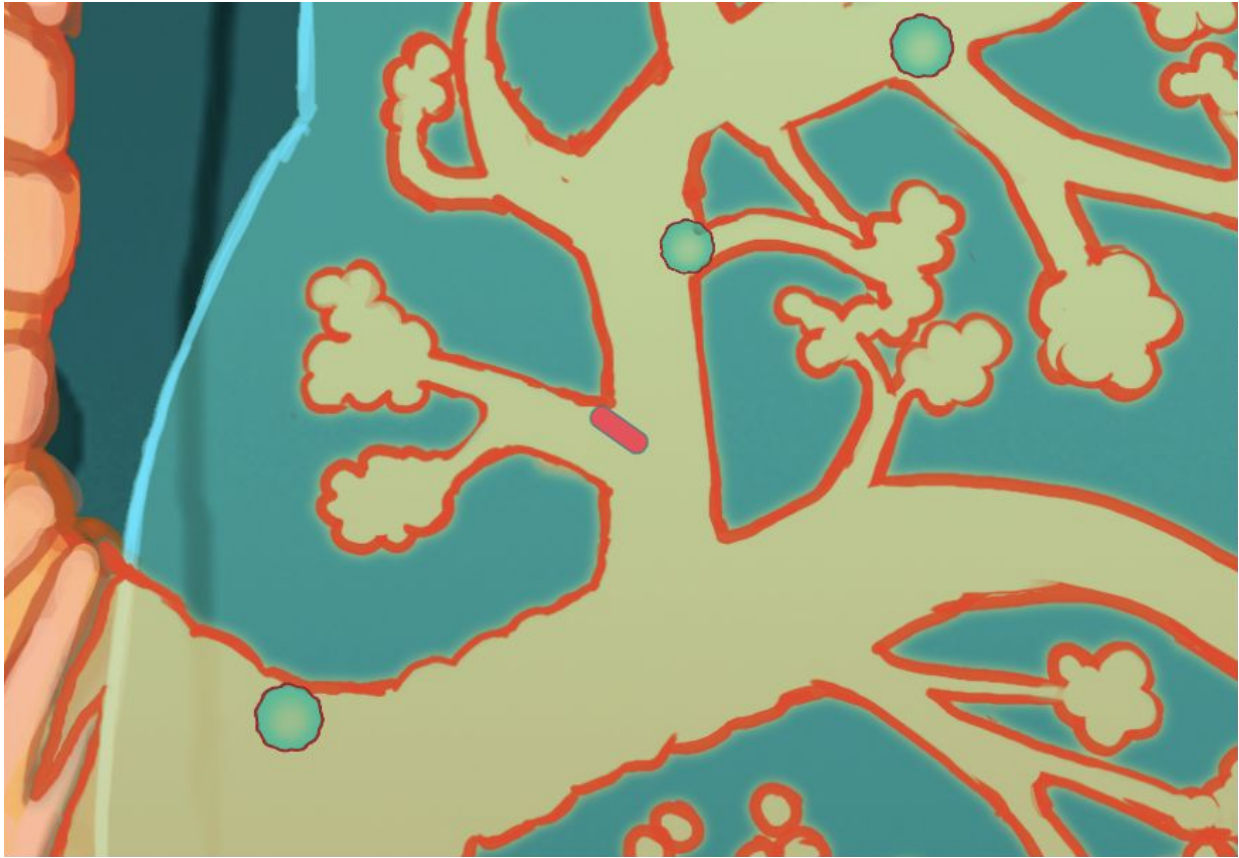

*Fig. 3.* The third level. The player (light-red bacilli) avoids enemies (circle-shaped white blood cells) in the bronchioles of the lungs.

*2.4. Assessments*

All assessments were digital-based as recommended by Habgood (2007) to conceal questionnaire length and facilitate marking and data analysis.

The performance assessment, used as the pretest and posttest, had 10 multiple choice questions, each with one correct answer, in a randomized order (Appendix A). Questions were chosen from Seeley, Stephens, and Tate's *Anatomy and Physiology* editions 6 and 7 (2003; 2004). The purpose of assessing performance is to see if *Grey Plague* was effective at promoting learning, one of *Grey Plague*'s purposes.

An engagement assessment for quantifying engagement with *Grey Plague* consisted of 25 questions using five-point Likert scales. Questions were selected from Jennett et al.'s (2008) Experiment 3 game immersion questionnaire. Seeing as immersion in a game is essentially a high level of engagement with the game (Brown & Cairns, 2004), the substitution seemed reasonable. Moreover, a questionnaire on immersion instead of engagement was used because little substantiated game engagement questionnaires were found. For example, O'Brien and Toms's (2010) User Engagement Scale, a self-report instrument that measures engagement with technology, has been frequently validated but does not target games specifically, making it less accurate (O'Brien & Tomes, 2013; Wiebe, Lamb, Hardy, & Sharek, 2014). Proposed self-report instruments that target engagement with games (e.g., Hamari et al., 2016; Wiebe, Lamb, Hardy, & Sharek, 2014) are less substantiated than Jennett et al.'s (2008) measurement, perhaps due to their recency. Selected questions were engagement-specific, addressing such factors as motivation, enjoyment, attention, effort, and other attitudinal, behavioral, and cognitive factors (Fredricks, Filsecker, & Lawson, 2016).

*2.5. Procedure*

Data for each participant were gathered on campus after classes. Each participant played on a MacBook Air with stereo sound through headphones.

Pretest score was collected, then observations were recorded as the participant played *Grey Plague* in approximately 30 minutes. Behaviors, body language, and facial expressions when interacting with the game and external environment were recorded. During this time, the researcher did not initiate communication with the participant.

Afterward, as feelings of engagement are most susceptible to memory decay, the engagement score was collected first (Gass & Mackey, 2016). Its Likert scales were treated as interval data, as suggested by Brown (2011), so that descriptive statistics could be applied.

Then, interview data were collected before collecting the posttest score to mitigate reactivity (Gass & Mackey, 2016). The face-to-face guided interviews were recorded with an audio device (Appendix B). Interview questions were modelled after SurveyMonkey's online example product feedback surveys (SurveyMonkey, n.d.). As suggested by Brenner (2006), the first few questions were descriptive, relatable, and broad to facilitate the interviewee's comfort with the questioning. Questions on game design, usability, and marketing were asked. Afterwards, the posttest score was gathered.

*2.4.1. Data analysis*

Following Creswell's (2007) and Yin's (2003) case study method, a within-case analysis was run to find unique themes for each case. Cross-case analysis then categorized similarities and differences between within-case findings.

Analyses were performed using NVivo 11.4.0, a qualitative analysis program. Following Wong's (2008) guide, an initial round of open coding to identify themes within each case resulted in 49 descriptive codes. Categories were developed by repeatedly removing overlapping categories, creating associations between categories, and capturing the main concept of remaining categories. Through this coding process, 10 game elements, categorized into 3 constructs, emerged.

### 3. Results and discussion

*3.1. Quantitative results*

Statistical analyses used SPSS 24.0 and a confidence level of 95%. The means of the dependent variables pretest, posttest, and performance (post-pre) within the groups are shown in Table 1.

**Table 1**

Descriptive statistics for pretest, posttest, performance, and engagement by group

| Variable | Group A | | Group B | |
|---|---|---|---|---|
| | M | SD | M | SD |
| Pretest[a] | 5.40 | 2.51 | 4.80 | 1.79 |
| Posttest[a] | 6.80 | 2.78 | 7.20 | 1.30 |
| Performance | 1.40 | 1.82 | 2.40 | 1.95 |
| Engagement[b] | 85.20 | 10.52 | 92.20 | 7.89 |

[a]The full score was 10.

[b]The full score was 115.

Preliminary analyses using Shapiro-Wilk test and Levene's test ensured compliance of parametric assumptions. A one-way MANOVA indicated no significant difference between groups on engagement or performance ($F(2, 8) = 0.79$, $p=0.49$). Within the multivariate analysis, univariate tests showed a stronger effect on engagement ($F(1, 9) = 1.42$, $p = 0.27$) than on performance ($F(1, 9) = 0.70$, $p = 0.43$). These results indicate that game B may have been more engaging than game A, thus conscious design decisions for game B might have been effective. The non-significance of these results may plausibly be because the sample size was too small, the playtime for the differentiated section between game A and game B was too short, and possibly the design differences between game A and game B were not drastic enough. These results implicate that the objective of differentiation between games - to gain diversified perspectives on the same topic - may not have been fulfilled.

A dependent *t*-test indicated a statistically significant improvement in test scores following the game treatment across groups from 5.10±2.08 to 7.00±2.06 ($t(9)=-3.24$, p=0.01): an improvement of 1.90±1.85. The effect size (d=1.90/1.85=-1.03) was calculated to be between 'large' and 'very large' (Cohen, 1988; Sawilowsky, 2009). Pretest (M1) and posttest ($M_2$) variables showed a positive correlation (r=0.59, n=10, p=0.07), substantiating the uniformity of the treatment effect across individuals. Given that the performance assessment was accurate, quantitative analyses indicate that *Grey Plague* is a valid educational game because it fulfilled the tacitly assumed goal of educational games: to have a significant impact on learning (Papastergiou, 2009; Ke, 2008; Annetta, Minogue, Holmes, & Cheng, 2009; Giannakos, 2013). Therefore, engaging game design suggestions for *Grey Plague* made by participants are rendered more applicable to effective educational games.

*3.2. Quantitative results*

Ten educational game elements for promoting engagement emerged from qualitative analyses. They are categorized into three different constructs - story, gameplay, and atmosphere - because these three constructs were the most frequently coded nodes during cross-case analysis. While all reported elements find common ground with entertainment game design, some aspects of these elements are unique to educational games because they additionally promote learning.

This section is first introduced by an overview of the identified elements (Table 2), then followed by discussions of each element.

**Table 2**

Engaging educational game elements, accompanied by description and example

| Game element | | Description | Example |
| --- | --- | --- | --- |
| Story | Tone | General attitude of the text, voice, or other communication that conveys the story. Forms 3D characters; facilitates empathy and understanding. Should be familiar or relatable. | "I think because a lot of the dialogue was very conversational or more like a stream of consciousness, I could understand it. My thought process is similar and the way I think things is very similar so it was easy to follow." (Sophia[1], on *Grey Plague*) |
| | Detail | Should have detail or clarity about game world, especially unique points about protagonist, setting, and learning content. | "I'd prefer a little more backstory.. I'm curious, I want more info about what [the protagonist's] childhood was like, and how she got her abilities." (Ava, on *Grey Plague*) |
| | Relationships | Interpersonal relationships developed in the game's societies. Should allow development of complex relationships, moral dilemmas, and drama. | "The relationship between [the protagonist] and mom and between [the protagonist] and grandma both served as an interesting contrast.. The whole idea of [the protagonist] being an outcast because of her strange ability was interesting." (Aiden, on *Grey Plague*) |
| | Fantastical reality | Common preference for reality-based stories, perhaps with fantastical twist. | "I want more realistic storylines instead of the typical post-apocalyptic future or robots taking over the world type of story; it's more relatable." (Ava, on games) |

---

[1]Pseudonyms are used to replace the names of participants.

| Game play | Clear goals and rules | Game goals and rules should be clear to players. Serves to avoid confusion or bad frustration and retain flow experience. | "Some parts of the game I couldn't really figure out what to do.. I didn't know you could interact with the people." (Evan, on *Grey Plague*) |
|---|---|---|---|
| | Identity | Ability to feel present as a unique individual in-game. Facilitated by allowing player-controlled character movement and customization. | ""I like the more immersive quality – you're the person in the game; you could act out the person and character.. How about a dressable character?" (Felicia, on *Grey Plague*) |
| | Experimentation/ exploration | Should have capability to autonomously explore game world, and experiment with game mechanics in game world. | "There were minor points in annoyance- like in the lungs, I kept on trying to go somewhere, and it was like 'no! You can't go there!'" (Felicia, on *Grey Plague*) |
| | Agency/autonomy | Ability to make meaningful decisions and express will, uninfluenced by extrinsic rewards, in-game. | "I felt limited in my choices, like I could only hit certain options, and I could only go to certain places and talk to certain people." (Katie, on *Grey Plague*) |
| Atmosphere | Visuals | i.e. interface, images, animations. Fosters unique atmosphere via style and color palette. Should have clear distinctions between shapes and natural character and environment animations. | "The people created in particularly the Sims 2 and 3 games seem really stiff in gesture and don't have enough interaction, making them a lot less appealing than what's shown on the trailers." (Hannah, on *Sims 2 & 3*) |

| | | |
|---|---|---|
| Audio | i.e. music, sound effects. Should be apropos to game world and event player is experiencing, and should avoid intrusive repetition. | "The audio was good. All the sound effects were really good. It was just really repetitive; after listening to it for about twenty minutes, I started to get kind of annoyed." (Katie, on *Grey Plague*) |

*3.2.1. Story*

According to participants and game researchers (Brown & Cairns, 2004; Ermi & Märyä, 2005; Schell, 2014), story is the series of events that the player experiences while playing; these events help shape the game world. Story can take multiple forms: linear or prescript or procedural; simple or complex (Schell, 2014). Participants found story at least as important as gameplay, if not more so, in attracting their initial interest in a game. Story elements, more than gameplay elements, have educational-game specific aspects. Engaging game story elements include interesting tone, detail, relationships, and fantastical reality.

*3.2.1.1. Tone*

Tone is the general attitude of the text, voice, or other communication that conveys the story. For example, tone could be humorous, serious, detached, introspective, upbeat, and so on. Tone is a well-known and fundamental story element in creative writing as well as digital storytelling (Ohler, 2013). For educational games, it helps form unique characters, facilitate empathy, and enhance understanding of learning content. In connection with participant interviews and usability and linguistic research (Nielsen, 1994; Ohler, 2013; Richardson, 2010), familiar or relatable tone is most effective in facilitating learning, while idiosyncratic tone, consistent but dependent on a character's mood, is best at fostering three-dimensional characters.

Dimensional, realistic characterization of non-player characters is important because it facilitates empathy and emotional impact and thus engagement (Brown & Cairns, 2004; Freeman, 2004). Additionally, the alternative - weakly written characters - is apparent and may be seen as amateur to the player, thus breaking immersion's suspension of disbelief if immersion was occurring (Freeman, 2004). Regarding tones for learning, one of Nielsen's (1994) ten usability heuristics for facilitating understanding of user interfaces dictates that dialogue should be expressed in language familiar to the consumer, and jargon should be avoided unless explained beforehand. Given this heuristic, unknown vocabulary thus should especially be explained in a player-friendly tone.

Note that tone would most likely not need to be considered for games with negligible communication.

*3.2.1.2. Detail*

Interviews indicated that sufficient clarity and detail about the game world, especially on the player-controlled character, setting, and the game world's connection to learning content, improved or would improve engagement and learning with games. For example, when asked "how could [*Grey Plague*'s story] be improved?", most participants wished for more detail on the protagonist's history, her powers, and the game world's setting: "more explanation of this future... more exposition" (Brenda, group B). Note that the desire for detail on setting might be exclusive to intriguing, non-realistic settings, e.g. futuristic; more research is needed to confirm.

A few participants felt that the learning and fiction components of *Grey Plague*'s story were disconnected; Katie, group A, recommended "incorporating more information into the storyline." This suggestion is in alignment with research on intrinsic integration (Habgood &

Ainsworth, 2011; Habgood et al., 2015). According to Habgood and Ainsworth (2011), educational games that employ intrinsic integration integrate the structure of the game world and core game mechanics with the learning content. This integration can then increase intrinsic motivation and learning performance (Habgood & Ainsworth, 2011). While some researchers (Habgood & Ainsworth, 2011; Lundgren & Björk, 2003) emphasize the importance of integrating the game mechanics over story with the learning content, this study would argue that the importance of integrating mechanics or game world and story (i.e. endogenous fantasy) is not consistent and largely dependent on the game in question. Indeed, educational games' aims for play are diverse (e.g. challenge, discovery, sensation) despite having a common purpose to teach (Aleven, Myers, Easterday, & Ogan, 2010). To evidence this claimed dependency on individual educational games, all complaints from group A about insufficient integration between *Grey Plague* and learning content were directed towards the story, plausibly because game A is mainly story-based and has no unconventional game mechanics. In any case, detail about the connection between learning content and the game world would be beneficial for the player's engagement as well as learning.

One novel consequence of insufficient detail, seemingly specific to educational games, is that it can beget misunderstandings about what is fiction and what is true. For example, three participants believed that the disease Grey Plague was real because they found the existence of mutated tuberculosis to be plausible or forgot Grey Plague took place in the future. This misunderstanding is an example of constructivist learning in a gaming context, whereupon new knowledge is constructed from previous knowledge (Bransford, Brown, & Cocking, 2000). Because new knowledge may be erroneous due to past misunderstandings or false beliefs (e.g.

Grey Plague is real because the game is assumed to occur in present time), any erroneous knowledge can thus be countered with sufficient detail via telling or guiding (Bransford, Brown, & Cocking, 2000).

*3.2.1.3. Relationships*

The player-controlled character - in this case, the protagonist - 's interpersonal relationships within the societies of the game world was engaging to participants; they found complexity, moral dilemmas, or drama especially interesting. For example, in *Grey Plague*, the protagonist faces a moral dilemma when they have the choice to divulge important, life-saving information at the expense of possibly giving away their secret about their superpower. Adrian, group B, wondered about the consequence of his choice in this situation carefully - "[the protagonist] probably wouldn't expose her powers, but I personally wanted to tell Joe." In connection with this element, Olsen (2010) finds strong social motivations to playing games, some of which include the motivation to compete, cooperate, or develop social bonds. Moreover, Przybylski, Rigby, and Ryan (2010) place large importance on this ability to connect socially within games, claiming that it satisfies the fundamental need for relatedness which, according to the self-determination theory (SDT), is necessary for intrinsic motivation (Ryan & Deci, 2000). Since intrinsic motivation positively affects engagement (Driscoll, 2005), social connection within games thus facilitates engagement. On top of that, Annetta (2010), in his framework for serious game design, lists social communication - called 'interactivity' - as one of six fundamental elements of an effective serious game, and elucidates the ubiquitous importance of interactivity across game modes (e.g. single-player, multiplayer) and multimedia environments.

Therefore, because substantial evidence exists to corroborate the engagement benefits of social communications, this game element on social relationships is validated as engaging.

*3.2.1.4. Fantastical reality*

Despite the game industry being inundated with games with unrealistic, oftentimes cliched stories (Schell, 2014), participants desired more realistic stories with a fantastical edge. As Ava, group B, recommended, "a story about some unfortunate soul who gets caught up in some shady business sounds great. I'd enjoy a semi-realistic story. Realistic with a twist to it." This element does not seem integral to improving story because only a few participants mentioned it, but this element could help the story appeal better to consumers. In fact, Chang et al. (2015), using statistical analyses, found that most students preferred a mix of fantasy and reality in serious games, with more older or female students preferring realistic scenarios over a post-apocalyptic fantasy scenario. Chang et al. concluded that serious games must have a balance between real-life and fantasy in order to optimize engagement. While their study limits student responses to at most five options for serious game story, their study is one of the only ones statistically examining student preferences for story in an educational game. More thorough investigation into student game story preferences would help ascertain or deny the validity of this game element.

A reality-based story has the added, potential benefit of facilitating learning by placing learning content in an authentic context. Authentic contexts, also known as authentic activities, are contexts for an educational task that resemble the context of an actual practitioner of that task (Cordova & Lepper, 1996). Education researchers (e.g., Cordova & Lepper, 1996; Light & Butterworth, 2016) use 'situated learning' to describe learning via authentic contexts. These researchers propose that situated learning increases learning and transfer of new knowledge. With *Grey Plague*, most participants cited situated learning as the cause for facilitating their understanding and appreciation of anatomy and physiology:

- Samuel, group B: The game definitely conveys information better... It gave it context, actual application.
- Ava: My interest increased because I could imagine 3D-printed organs being mass produced and helping people.

This finding on the efficacy of reality-based stories can connect to Habgood and Ainsworth's (2011) and Habgood et al.'s (2015) findings that intrinsically integrated educational games facilitate engagement and learning better than non-integrated ones. Conclusively, fantastical, reality-based stories can increase affective engagement and, if integrated with learning content, facilitate learning; however, player preferences should be taken into account as they might affect efficacy.

*3.2.2. Gameplay*

By employing previous definitions of gameplay in the gaming literature (Brown & Cairns, 2004; Ermi & Märyä, 2005; Schell, 2014), this study defines gameplay as the player's interactions in the game and the impact of those interactions. Identified, engaging gameplay elements include clear goals and rules, identity, experimentation and exploration, and agency and autonomy.

*3.2.2.1. Clear goals and rules*

Goals are what the player should achieve to progress in the story; rules define what the player and other entities can do. This element dictates that the player should be certain about the goals and rules of an educational game. When uncertainty about game goals or rules occurs, the player becomes frustrated in-game and thus less engaged. Germane cognitive load is redistributed to attend to resolving this uncertainty; when the player's attention is on discovering the goal and not the goal itself, the player could make false discoveries or deductions which will further decrease engagement (Osman, 2010).

Goals and rules are amongst the oldest and most fundamental components of games (Garris, Ahlers, & Driskell, 2002; Rieber, 1996). Moreover, the importance of this game element has consistently been reiterated across many different disciplines throughout the past decades (Driskell & Dwyer, 1984; Ricci, Salas, & Cannon-Bowers, 1996; Garris, Ahlers, & Driskell, 2002; Federoff, 2002; Pinelle et al., 2008; Schell, 2014; Sweetser & Wyeth, 2005). As a result of this consistent and prolific literature, only relatively novel aspects of this element are reported.

Some causes for uncertainty about goals and rules include giving the player new in-game abilities (e.g. talking to people, getting and using items) without notice or directions about their newfound powers, or contradicting the player's assumptions about meaningful objects. This concept of 'meaning' seems to be novel to the gaming literature. To clarify, meaning is an attribute of an object and signals that its object will facilitate advancement in the game. Players typically attribute meaning to locations, non-player characters, and items. Ancillary content such as the floor or sky will be ignored as meaningless unless the player knows otherwise. When aforementioned, assumed attributions are contradicted (i.e. when something assumed meaningless is meaningful and vice versa), players become frustrated thus unengaged. For example, two out of three stores in *Grey Plague* are meaningless to game progress, but participants intuitively attributed meaning to them. This led to many participants repeatedly returning to the stores, believing these stores contained the solution to their goal. As Felicia, group B, noted, "I was struggling... I figured I could click on [the store] for a reason." In alignment with this reasoning, Federoff (2002) stated that, according to Nielsen's (1994) usability heuristic #2, elements in the game world should function similarly as they would in the real world. Moreover, Pinelle et al. (2008) identified two game usability heuristics which find that responses to player's actions should be consistent and predictable. Thus, goals and rules that contradict assumptions about meaning should be consistent and predictable or clarified to the player.

Clear goals can be established via making relevant information available to the player (e.g. a virtual to-do list or quest list). This method allows the player's cognitive load reprisal so it can attend to other events; additionally, negative effects from worrying over remembering goals will be mitigated (Nielsen, 1994; Pinelle et al., 2008).

For rules, intuitive or conventional rules (e.g. type the protagonist's name in the input field) were discovered by participants via trial and error and thus needed no guidance. Similarly, Pinelle et al. (2008) finds that goals and rules could be intuitively deducted through trial and error, with clear, consistent hints for slower learners. However, counterintuitive rules require explicit, well-worded directions. Directions must be well-written - in *Grey Plague*, confusion was engendered two times by poor wording: "I just misread something as 'right click' instead of 'click to the right'" (Edward, group A). Similarly, Nielsen's (1994) usability principle states that directions should be in precise, plain language and answer any confusion, and Pinelle et al.'s (2008)'s heuristic asserts that complex games should provide instructions. Moreover, Desurvire,

Caplan, and Toth (2004) state that instructions should mimic gameplay, stay within game world and not break suspension.

Because educational games have a learning component, the implementation of this game element is more complicated for educational games than entertainment games (Annetta, 2010). For instance, one participant who didn't comprehend a section of the learning content was uncertain about goals and rules requiring knowledge of said content. Comprehension of learning content may be needed to establish certainty; perhaps more research is needed to affirm.

*3.2.2.2. Identity*

Identity is the player's ability to feel like a unique individual in the game. In accordance with four participants and psychology research (e.g., Sedikides & Brewer, 2015), identity appeals to the psychological human trait to form a self-concept, or a collection of beliefs, about oneself, which then in turn affects one's learning, affect, self-esteem, and life-satisfaction. For example, Sophia, group A, commented, "because I was acting as some character in a story.. It made it more interesting to me, more relatable." Specifically, the game element 'identity' satisfies one of three fundamental self-representations that comprise self-concept: the individual self. Theoretically, following the research on individual self, optimal identity would then allow differentiation from other character in order to enhance the player psychologically and thus maximize engagement (Sedikides & Brewer, 2015).

For more concrete methods of implementing identity, perhaps existing research can shed some light. Identity is, similar to clear goals and rules, a commonly perused game element in gaming and educational gaming research (Annetta, 2010; Gee, 2003; Gee, 2005; Ke, 2008; Squire, 2006). Some of these researchers refer to an 'avatar', or a visible player-controlled character, when describing how to evoke identity. The player projects their mental state and autonomy into this avatar; this avatar then communicates this state and autonomy as interaction or other information to other characters (Annetta, 2010; Schell, 2014). Annetta and Holmes (2006) found that differentiation from other characters, necessary for effective identity, was enabled via increased avatar options. Consequently, a game with 100 unique avatars was found more engaging than a game with only 2 (Annetta & Holmes, 2006). In a more recent, rigorous, and statistically sound study, Turkay and Kinzer (2014) found that avatar-based customization positively impacted players' identification with their avatars. Furthermore, when asked "What are the things that you would like to improve in this game?", one participant enthusiastically asked, "How about a dressable character? Dress-up!" Therefore, this study would propose character customization as one concrete method for facilitating identity.

Another method for identity could be player-controlled character movement. For instance, six participants expressed desire for more identity through controlling the protagonist's movement in space: "You could move her going somewhere" (Ava) or "If it wasn't just text based, and you could direct the character, that would be good" (Brenda). There is an absence of research on the effect of player-controlled movement on identity or engagement. Perhaps this absence is because games are automatically assumed to have player-controlled movement (e.g., Reynold, 1999).

*3.2.2.3. Experimentation and exploration*

Participants, especially those from group B, were observed to be engaged in autonomously experimenting with the game mechanics and exploring the game world, even if these actions were meaningless to their progress. Restrictions to experimentation or exploration should have an explanation, or else players might become confused and uncertain. Felicia, when regarding an area where her character could not go, remarked that "even though I was kind of annoyed at that part, it was for a good cause. It told me why it was better to go in the upper half [of the level]."

This game element has connections to existing entertainment gaming research. For instance, Federoff (2002) notes that experimentation and exploration stem from the player's desire for control and freedom. Given this reasoning, this game element can be induced by the following: provision of feedback after 0.2 to 0.4 seconds in response to the player's control input (Bickford, 1997; Kimble, 1947), customizable control mappings (Pinelle, Wong, & Stach, 2008), ability to save and load game states, and minimal exploration restrictions (Federoff, 2002). These listed methods will provide feelings of control and thus confidence and intrinsic motivation to explore or experiment (Federoff, 2002). Perhaps the desire for user control and freedom stems from a broader psychological need for autonomy, as defined in SDT, which would then enable growth of intrinsic motivation (Ryan & Deci, 2000). Indeed, cognitive evaluation theory (CET), a subtheory of SDT, predicts that intrinsic motivation and exploration are positively correlated (Ryan & Deci, 2000). Additionally, Martens, Gulikers, & Bastiaens (2004) found that students with more intrinsic motivation tended to explore more of an instructional technology to satisfy their curiosity. Note that intrinsic motivation is dependent on conscientiousness, one of the personality traits listed in the 'Big Five' which is characterized by self-control, diligence, organization, detail-mindedness, and responsibility, among others (Barrick & Mount, 1991; Clark & Schroth, 2010).

This experimenting, exploring behavior additionally aligns with the philosophies of exploratory learning and self-directed learning. Self-directed learning, or learning that is started and managed by the learner, has gained recent academic attention, especially on its use with instructional technology (e.g., Kim, 2010). In fact, Rashid and Asghar (2016) found that technology use directly, positively affected self-directing learning.

Conclusively, this engaging game element only occurs if the player's psychological needs are satisfied and the player is then intrinsically motivated; additionally, this element is mediated by the player's conscientiousness. Upon its occurrence, this element can then facilitate engagement and self-directed learning.

*3.2.2.4. Agency and autonomy*

Agency is the ability to make meaningful decisions in a game. In accordance with Rigby and Ryan (2011) and Ryan and Deci (2000), autonomy is the player's psychological need to express their will, uninfluenced by extrinsic rewards. Participants mentioned both as important factors for gameplay. For example, Hannah, group A, commented, "I really like games where you have the option of choosing what you want to do.. I like games like Minecraft, where you can build stuff, while at the same time you have the option of interacting with other players or even making your own art or killing monsters."

To improve agency and autonomy, participants suggested a greater quantity of meaningful choices paced evenly throughout the game. Participants were most often dissatisfied with a lack of choices in *Grey Plague*.

*3.2.3. Atmosphere*

Brown and Cairns (2004) define atmosphere as the unique mood defining the game world, composed of constantly attended visuals and audio. Brown and Cairns also point out that it is the barrier to immersion; i.e. it is required for immersion. The focus of atmosphere on visuals and audio aligns with the tendency for educational gaming and cognitive research to address visuals and audio in tandem, perhaps with 'audiovisual' or 'aesthetics' (Annetta, 2010; Ermi & Mäyrä, 2005; Mayer & Moreno, 2003). In congruity with this focus, Aiden, group B, said, "I took in the art and music of *Grey Plague*... They created a surreal, dreamlike atmosphere."

Interviews agreed with Brown and Cairns that atmosphere is composed of visuals and audio; however, in contrast with Brown and Cairns's claims, interview data did not signal that visuals and audio must be constantly attended to in order to facilitate immersion: "the audio.. I could not remember if it changed because I associated it with the place. Thought it fit" (Samuel). Further evidence (Annetta, 2010; Sanders & Cairns, 2010; Wickens, 1992) demonstrates that unless audio is abrupt, auditory signals will typically only be processed at the preattentive (i.e. subconscious) level; that more immersive audio does not necessarily need to be more complex or take up more cognitive processing; and that, while games tend to dominate a player's cognitive channel for processing visuals rather than audio, this "visual dominance" (Annetta, 2010, p. 108) is largely dependent on the goals of the game. This study reasons that, in conflict with Brown and Cairns, immersive visuals and audio do not constantly need to be processed at the attentive level; although, visuals seem to be attentively processed more than audio.

Despite this conflict, interviews and previous research (Ermi & Mäyrä, 2005; Jennett et al., 2008; Klimmt, 2003) do indeed agree that atmosphere enables immersion; thus, atmosphere's concomitant components, visual and audio, are not required to evoke lower levels of engagement. If an immersive, highly engaging educational game experience is desired, effective atmosphere should be implemented.

*3.2.3.1. Visuals*

The visuals of a game refer to the real-time display of content which is visible in the player's visual field (Jennett et al., 2008); visuals are then processed in the visual channel of the human information-processing system (Mayer & Moreno, 2003). Player attention to visuals tends to follow a pattern: upon initial engagement, attention is widely distributed over the visual field; upon higher levels of engagement, attention is narrowed and fixed (Jennett et al., 2008; Styles, 2006). Previous studies (Chang, Liang, Chou, & Lin, 2017; Chen & Sun, 2012; Mayer & Moreno, 2003) have found that high quality, dynamic visuals (e.g. video, animation) of a cognitively manageable quantity and pace are most effective in increasing engagement.

The two most engaging aspects of visuals found are concurrently sparse the gaming research: clarity and natural animations. Clarity is important because it facilitates the player's understanding of their surroundings or context and decreases confusion or frustration. For instance, Brenda, group B, desired clarity in one of the static visuals because "the girl seemed to melt into the blankets.. It was kind of weird." Likewise, Spence and Feng (2010) found that players expect clarity in their attentional visual field because the smallest details could influence the player's outcome. Clarity can be achieved through design principles such as shapes, contrast, and edges.

Animations, such as character or environment animations, should feel natural. Characters' velocities and animations should comply with the laws of physics. Movement that seems unnatural is received negatively; participants referred to poor movement in games as "clunky" or "annoying." Poor clarity or animations can cause the atmosphere to seem artificial. If the atmosphere feels fake, the suspension of reality and disbelief that characterizes immersion will be broken (Brown & Cairns, 2004; Dede, 2009; Ermi & Mäyrä, 2011). Existing research on natural animations is largely technical and focused on 3D animation (e.g., Aguiar, Zayer, Theobalt, & Seidel, 2007; Welbergen et al. 2010). The role of animation in games has not been specifically studied, perhaps because there is a general lack of scholarly study on animation. Nonetheless, clarity and natural animations help make the game world more believable by breathing life into it and letting players better decipher it.

*3.2.3.2. Audio*

Audio can be defined as auditory input that is decoded, encoded, then processed in the auditory channel of the human information-processing system (Mayer & Moreno, 2003). Audio helps the listener detect danger, navigate, and communicate (Burton, 2015). Game audio typically consists of music, sound effects, and voiceovers or other verbal communications (Marks, 2013). Currently, game audio literature (Marks, 2013; Sanders & Cairns, 2010; Stevens & Raybould, 2013) has found that audio is immersive when it is positively perceived by players, meets expectations and conventions of game audio, and is not intrusively repetitive. Interview data identified two immersive aspects of game audio, both which are noted in existing literature: conventions, and lack of intrusive repetition.

Conventional audio essentially follows and adapts to the player-controlled character's mood and situation. As Hannah explained it, "if you find accomplishment, the music should be uplifting.. the other way around, the music is gloomy as hell." For instance, conventional music might include main-menu or introduction music, background music, "win" or "lose" music, and closing music (Marks, 2013, p. 187-192). Conventional sound effects might include footsteps for character movements, ambient noises, sound effects for collisions, and sound effects associated with certain game mechanics, objects, or situations (Friberg & Gärdenfors, 2004; Marks, 2013; Stevens & Raybould, 2013). Additionally, conventional techniques from other industries, such as film, can additionally facilitate engagement (Whalen, 2004).

To prevent looped or repeated audio from gaining unnecessary, negative attention to the repetition, audio should contain variety, with the amount of variety dependent on the educational game. Katie noted her dislike of repetition; "The music.. was just really repetitive; after listening to it for about twenty minutes, I started to get kind of annoyed." According to Stevens and Raybould (2013), the amount of variety necessary is determined by a game's structure, especially the game's playtime length and genre. Stevens and Raybould elucidate how game genre can affect optimum variety: casual or puzzle games require minimal variety because players have low expectations and often will play with the game audio turned off; retro or arcade games need little variety because players expect a repetitive but catchy, non-intrusive tune; simulation games, such as driving or sports games, also need little variety due to their repetitive game mechanics; role-playing or adventure games need moderate variety due to their long playtime; and so on so forth. Game audio literature (Marks, 2013; Stevens & Raybould, 2013) finds generative and procedural techniques effective for saving data storage space while implementing audio variety. Specifically, variety in sound effects can be executed via automated, randomized adjustments to the frequency, timbre, volume, or other parameters. For variety in music, track length can be increased to avoid excessive looping. However, a more economic, procedural technique would be to load the elements of the music (e.g., bass, harmony, melody) into the game and automatically randomize their parameters during real-time play. For example, to use this economic method in Unity, each audio component can be put into an Audio Mixer Group, whose parameters can then be accessed and randomized while the game runs.

## 4. Conclusions

In closing, to help fill the gap on engaging educational game design elements, this study identified ten elements for promoting engagement in educational games, and categorized the identified elements into three common constructs (i.e. the three most coded top-level 'nodes'): story, gameplay, and atmosphere. Within story, the following engaging elements were found: (1) tone, the general attitude of the communication conveying the story, which additionally facilitates learning; (2) detail about the game world, which can clarify educational concepts; (3) relationships within the game societies; (3) and fantastical reality-based story themes, which can facilitate situated learning but is dependent on player preference. Within gameplay, engaging elements were identified: (1) clear goals and rules; (2) identity; (3) experimentation and exploration, which can facilitate learning if directed towards learning purposes; and (4) agency and autonomy. Within atmosphere, the barrier to immersion and most engaging construct, (1) visuals were found to be immersive if clarity and natural animations are provided; (2) audio was immersive if it complied with certain conventions and avoided intrusive repetition. Overall, game design to promote engagement within educational games versus entertainment games has some differentiation. Although all ten identified elements find connections in the usability and gaming literature, four out of the ten elements, most in the story construct, were definitively unique because they were complicated by the learning aspect of educational games.

These findings have some novel implications that will hence be discussed. Firstly, a majority of the novel findings are composed of specific methods to induce identified elements, which implicates that the research on educational game design technique over theory is sparse. Moreover, a majority of existing literature on game design technique is comprised of books that, while they come from credible or experienced authors, may not come from academic publishers. To benefit educational game developers, perhaps more technical educational game design research is needed.

Secondly, the findings implicate that the game design of educational games should implement a diverse variety of learning philosophies (e.g. situated learning, exploratory learning) in order to accomplish teaching as well as engagement purposes. However, if only the promotion of engagement is considered, educational games can plausibly employ entertainment game design principles and heuristics with success, and can additionally facilitate learning with certain elements. Thus, the gap in engaging game elements for educational games may have already been partially filled by previous game design literature (Bethke, 2003; Pinelle, Wong, & Stach, 2008; Rigby & Ryan, 2011; Schell, 2014).

This study opens new avenues for future work. Similar future research should include larger samples, greater differences between game versions, and more extensive, thorough testing. Additionally, participants should be asked about or debriefed on the definition of unfamiliar terms in interviews, such as "story" and "gameplay," in order to increase validity.

Future fruitful areas of research include educational game design that targets specific characteristics or attitudinal factors, or effective educational game marketing. Aforementioned topics could facilitate educational game development and use.

*4.1. Limitations*

This multiple case study examines a small sample of participants who, despite proportional diversity in gender, race, academic achievement, and gaming habits, were from the same school in a high-income county. Regarding data collection, engagement and interview data was self-reported, which can be unreliable due to probability that participants are unable or unwilling to report actuality. Additionally, in hindsight, the performance assessment questionnaire was not optimal due to its limited number of questions and multiple choice nature, which allowed falsely correct answers via chance. Caution should be exercised when generalizing findings to other instructional technologies and applying findings to populations with different demographics.

Appendix A

After air leaves the trachea, it goes through the:

Bronchioles

Bronchi

Alveoli

Capillaries

Gaseous exchange takes place in the lungs in the:

Bronchioles

Bronchi

Alveoli

Cardiac muscle is:

found everywhere

located in the abdomen

unique to the heart

(Select all that apply.) Select all of the following that are symptom(s) of tuberculosis.

Cavitation

Granulomas

Nodosomes

Clear sputum

Tuberculosis is transmitted through:

Infected water

Infected hands

Infected blood

Infected air

Which is most commonly collected to diagnose respiratory infections?

Saliva

Breath

Sputum

Any of the above

None of the above

What do Mycobacterium tuberculosis have that prevent their destruction by the immune system?

Cord factor

Exotoxins

Special protein

Capsule

Endotoxin

Which of the following is NOT one of the lobes of the cerebral hemisphere?

Ethmoid

Frontal

Occipital

Temporal

The central nervous system consists only of the brain.

True

False

Which of the following statements best describes homeostasis?

Keeping the body in a fixed and unaltered state

Maintaining a balanced internal environment

Altering the external environment to accommodate the body's needs

Appendix B

1. How have your opinions of Anatomy and Physiology changed after playing the game? For example, a changed interest in the topic? Any changed beliefs?
2. After playing the game, would you like to get involved more with Anatomy & Physiology, for example learn more about anatomy or medical sciences in your free time, and why or why not?
3. How much did you trust the information about anatomy given in the game?
4. Did you think Grey Plague was a real disease?
5. How do you feel about learning about anatomy in a game?
6. What do you think you learned from the game?
7. What problems with the game did you run into?
8. Did you get confused/annoyed by the controls at any point in time?
9. Would you buy the finished game at your chosen price? **If no,** why not? **If yes,** why, and how much would you pay for it?
10. During the first few minutes of playing, what was your initial opinion of this game?
11. After playing the game, did your opinion change, and if so, how?
12. How do you learn about games? E.g. through social media, news, word of mouth, etc?
13. When sharing a game to others, what do you use to share? E.g. email, facebook, twitter, etc?
14. When learning about a game, what information do you like to see provided? Information about the story, or about the unique features, etc?
15. How entertaining was the game?

16. How were the visuals? How could it be improved?

17. The story? How could it be improved?

18. The gameplay? How could it be improved?

19. The audio??

20. What aspect of games are most important to you—visuals, story, gameplay, etc?

21. What are the things that you would like to improve in this game?

22. What do you like most about this game?

23. What would you most like to improve/change about edutainment in general?

24. What would you most like to improve/change about games in general?